\documentclass{pos}

\title{Single pion production induced by neutrino-nucleon interactions}

\ShortTitle{Single pion production induced by neutrino-nucleon interactions}

\author{\speaker{Krzysztof M. GRACZYK}%
         \thanks{The author was supported by the grant:  35/N-T2K/2007/0 (the project number DWM/57/T2K/2007).
}\\
        Institute of Theoretical Physics, University of Wroc\l aw\\
        E-mail: \email{kgraczyk@ift.uni.wroc.pl}}

\abstract{This talk presents some of the results of the re-analysis \cite{Graczyk:2009qm} of the bubble chamber
          data for single pion production induced by neutrino scattering off deuteron.
          It is shown that ANL and BNL data are statistically consistent.
          The validity of the Adler relations (between $P_{33}(1232)$ resonance axial form factors) is also investigated.
          }

\FullConference{European Physical Society Europhysics Conference on High Energy Physics,
EPS-HEP 2009,\\
		 July 16 - 22 2009\\
		 Krakow, Poland}

\begin{document}

\section{Introduction}

Detailed experimental studies of the neutrino oscillation require more precise
knowledge of the cross sections for the interaction of neutrinos with matter.
The new long-baseline experiments, like T2K \cite{T2KLOI}, will be able to measure  neutrinos with higher  than past precision. Therefore, the theoretical and experimental studies of the neutrino scattering off nucleon/nucleus is of wide interest \cite{NuInt09}.

In the T2K experiment the neutrino beam energy has a peak at 0.7~GeV.
For such neutrino energies two types of interactions  are mainly observed: (i) quasi-elastic (charged current (CC) interactions) or elastic (neutral current (NC) interactions);
(ii) inelastic scattering, with $1\pi$ production. Both interactions are important  for the investigation of the $\nu_\mu\to\nu_\tau$ oscillation. While
the $1\pi^0$ production (by NC) is crucial for measurement of $\nu_\mu\to\nu_e$ oscillation\footnote{ $1\pi^0$ events constitute the background for the measurement of the electrons produced in $\nu_e A$ scattering.}.

In this talk we focus on the $1\pi$ production induced by neutrino-nucleon interaction.
Last years the subject has been intensively  studied theoretically
\cite{nonres,other_th}, and experimentally \cite{experiments}.

In modern experiments the neutrino-nucleus scattering is observed. The lack of knowledge of the axial structure of the nucleus induces, into the data analysis, the additional systematical uncertainty, which is difficult to control. Therefore, it is still interesting to look at the old neutrino-deuteron scattering data in order to extract information about the neutrino-nucleon interaction.

This short talk presents some of the results of the re-analysis \cite{Graczyk:2009qm} of the old single pion production data (the neutrino-deuteron scattering data) collected at two different bubble chamber experiments, which worked at Argonne National Laboratory (ANL) \cite{Radecky:1981fn} and
Brookhaven National Laboratory (BNL) \cite{Kitagaki:1990vs}.

\section{Re-analysis of the bubble chamber data}

The reaction $\nu_\mu + d \to \mu^- + \pi^+ + p + n$ is a subject of re-analysis. This is the simplest channel for $1\pi$ production, because pions are produced mainly by  excitation of the nucleon to $P_{33}(1232)$ resonance and  the nonresonant contribution seems to be negligible. However, there are theoretical approaches which contain also the non-resonant dynamic in the description \cite{nonres} of this channel.

In our analysis we assumed that the nonresonant contribution can be neglected, while the excitation of the nucleon to $P_{33}(1232)$ resonance is described by the hadronic current, which is expressed in terms of vector and axial form factors. The vector form factors are constrained by the electroproduction data (by CVC theorem). Therefore,  only
the axial form factors have to be established by the neutrino scattering data. The axial current is expressed in terms of four form factors: $C_3^A(Q^2)$, $C_4^A(Q^2)$, $C_5^A(Q^2)$, and $C_6^A(Q^2)$. But
PCAC hypothesis relates $C_6^A(Q^2)$ with $C_5^A(Q^2)$. Additionally, to simplify the problem, the Adler relations are usually postulated:
\begin{equation}
C_3^A(Q^2)=0, \quad C_4^A(Q^2) = - C_5^A(Q^2)/4.
\label{adler_relation}
\end{equation}
Then, only $C_5^A$ axial form factor leaves to fit to the ANL and BNL data.

\begin{figure}[t!]
\centering{
\includegraphics[width=1.0\textwidth, height=8cm]{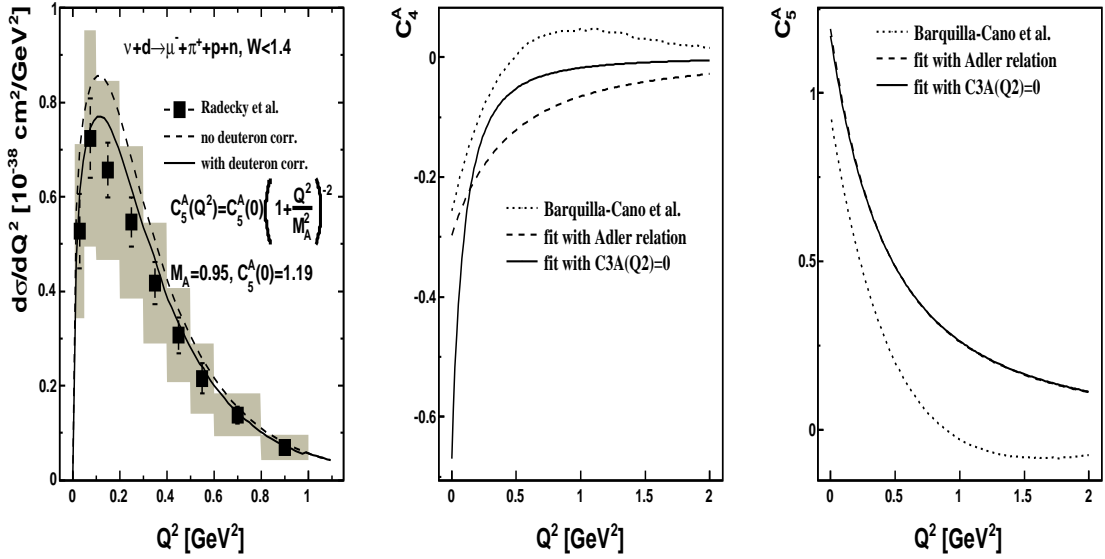}}
\caption{In the first figure the $d\sigma/dQ^2$ cross sections computed for the ANL beam are plotted (data were taken from \cite{Radecky:1981fn}). The solid/dashed lines denote the cross sections reduced/not reduced by the deuteron correction. In the second and third figure the $C_4^A$ and $C_5^A$ form factors are plotted. The fits (3.2) are denoted by solid lines. The fits (2.2), computed with $C_4^A=-C_5^A/4$, are plotted with dashed lines. The quark model predictions from Ref. \cite{BarquillaCano:2007yk} are plotted with dotted lines.
\label{rysunek}
}
\end{figure}

Since our analysis deals with two independent experimental data sets, to get a reasonable global fit,  the additional systematical uncertainty (normalization error) is required to take into consideration\footnote{It is the standard statistical treatment of independent  data sets, see e.g. \cite{Alberico:2008sz}.}. It significantly increases the $d\sigma/dQ^2$ cross section data uncertainty. In Fig. \ref{rysunek} the total cross section error (sum of  statistical, not correlated systematical, and normalization errors) is plotted by the shadow area.

The deuteron structure correction is also taken into account (for details see \cite{Graczyk:2009qm}). It makes the discussion more complete. The deuteron effect turned out to be more important for the ANL data (see Fig. \ref{rysunek}, where $d\sigma/dQ^2$ cross sections with and without deuteron correction are plotted).

For a global fit of $C_5^A$ we analyze two different parametrizations (dipole and so-called Adler). The best fit was obtained for:
\begin{eqnarray}
\label{global_fit}
C_5^A(Q^2) &=& (1.19 \pm 0.08) \left(1 + Q^2/(0.94 \pm 0.03\,\mathrm{GeV})^2  \right)^{-2},\quad \mathrm{and}\\
C_5^A(Q^2) &=&  (1.14 \pm 0.08)\left(1 - 1.21 Q^2/(2 + Q^2)\right)
\left(1 + Q^2/(1.29 \pm 0.07\,\mathrm{GeV})^2  \right)^{-2}. \nonumber
\end{eqnarray}
Both fits were computed with goodness-of-fit (GoF) larger than 58\%. The normalization of data  was obtained as follows: 1.08 (for ANL), and 0.98 (for BNL). The statistical quality of the fit was positively verified  by applying parameter-goodness-of-fit test (see Ref. \cite{Graczyk:2009qm}). The test  showed that the ANL and BNL data are consistent. Eventually, the form factors (\ref{global_fit}) were implemented to NuWro Monte Carlo generator.
It is interesting to notice that the CC$1\pi^+$/CCQE cross section ratio, computed with NuWro for MiniBooNE experiment,     is in an agreement with experimental measurements \cite{MBRatio}.

%###########################################################
%#############################################################

\section{Validity of the Adler relations}

It seems interesting to investigate the validity of the Adler relations (\ref{adler_relation}), which were originally obtained from the dispersion theory.

First of all, let assume that the form factors $C_3^A$, $C_4^A$, $C_5^A$  are independent but to reduce the number
of unknown parameters we assumed that they are parameterized as follows:
\begin{equation}
 C_i^A(Q^2) = C_i^A(0) \left(1 + Q^2 /M_{A,i}^2  \right)^{-2}, \quad i=3,4,5.
\end{equation}
Thus the global fit consists of six form factor parameters. For such fit the estimated values of the parameters turned out to be strongly correlated, but $C_3^A$ form factor appeared to be very small (of order of $10^{-5}$) and consistent with first Adler constrain. Therefore, it is reasonable to reduce the number of fit parameters by setting $C_3^A=0$. It allows us to investigate more carefully second Adler relation.

With $C_3^A=0$ constrain the best fit is computed for:
\begin{equation}
 -C_4^A(0) = 0.67 \pm 0.42, \; M_{A4} = 0.4^{1.1}_{0.4}\, \mathrm{GeV}\quad
 C_5^A(0) = 1.17 \pm 0.13, \; M_{A5} = 0.95 \pm 0.07\,  \mathrm{GeV},
 \label{fit_c4a}
\end{equation}
where $\chi^2/NDF=23.7/26=0.91$  and GoF =$59\%$.  The form factors $C_4^A$ and $C_5^A$ are plotted in Fig. \ref{rysunek}. The fit (\ref{fit_c4a}) is compared with the previous one (\ref{global_fit}) (here $C_4^A=-C_5^A/4$) and quark model predictions of Ref. \cite{BarquillaCano:2007yk}. One can see that obtained results (\ref{fit_c4a}) do not exclude also the second Adler relation, however, the $C_4^A$ form factor is characterized by large uncertainty.
Therefore, to perform more detailed analysis of axial form factors, more precise experimental data, in a wide range of the  scattering angle, is required.

\end{document}